\documentclass[twocolumn,tighten]{aastex62}
\pdfoutput=1 %for arXiv submission
\usepackage{amsmath,amstext}
\usepackage[T1]{fontenc}
\usepackage{ae,aecompl}
\usepackage[utf8]{inputenc}
\usepackage{apjfonts} 
\usepackage[figure,figure*]{hypcap}
\usepackage{enumitem}
\usepackage{url}
\usepackage{bm}
\usepackage[normalem]{ulem}
\input{hyperlink-year-only-natbib-patch}

\graphicspath{{./}{figures/}}

\begin{document}

\title[Gwave]{
The binary--host connection: astrophysics of gravitational wave binaries from their host galaxy properties} 

\correspondingauthor{Susmita Adhikari}
\email{susmita@stanford.edu}
%\author[0000-0001-5417-2260]{Authors*}
\author[0000-0001-5417-2260]{Susmita Adhikari}
\affiliation{Kavli Institute for Particle Astrophysics and Cosmology and Department of Physics, Stanford University, Stanford, CA 94305, USA}
\author[0000-0002-1980-5293]{Maya Fishbach}
\affiliation{Department of Astronomy and Astrophysics, The University of
Chicago, Chicago, IL 60637, USA}
\affiliation{Kavli Institute for Cosmological Physics, The University of Chicago,
Chicago, IL 60637, USA}
\author[0000-0002-0175-5064]{Daniel E. Holz}
\affiliation{Department of Astronomy and Astrophysics, The University of
Chicago, Chicago, IL 60637, USA}
\affiliation{Kavli Institute for Cosmological Physics, The University of Chicago,
Chicago, IL 60637, USA}
\affiliation{Department of Physics, The University of Chicago,
Chicago, IL 60637, USA}
\affiliation{Enrico Fermi Institute, The University of Chicago,
Chicago, IL 60637, USA}
\author[0000-0003-2229-011X]{Risa H. Wechsler}
\affiliation{Kavli Institute for Particle Astrophysics and Cosmology and Department of Physics, Stanford University, Stanford, CA 94305, USA}
\affiliation{SLAC National Accelerator Laboratory, Menlo Park, CA 94025, USA}
\author{Zhanpei Fang}
\affiliation{Kavli Institute for Particle Astrophysics and Cosmology and Department of Physics, Stanford University, Stanford, CA 94305, USA}



\begin{abstract}

Gravitational waves produced from the merger of binary neutron stars (BNSs) are accompanied by electromagnetic counterparts, making it possible to identify the associated host galaxy. We explore how properties of the host galaxies relate to the astrophysical processes leading to the mergers. It is thought that the BNS merger rate within a galaxy at a given epoch depends primarily on the galaxy's star-formation history as well as the underlying merger time-delay distribution of the binary systems. The stellar history of a galaxy, meanwhile, depends on the cosmological evolution of the galaxy through time, and is tied to the growth of structure in the Universe. We study the hosts of BNS mergers in the context of structure formation by populating the Universe Machine simulations with gravitational-wave events~ according to a simple time-delay model. We find that different time-delay distributions predict different properties of the associated host galaxies, including the distributions of stellar mass, star-formation rate, halo mass, and local and large-scale clustering of hosts. BNSs that merge today with short delay times prefer to be in hosts that have high star-formation rates, while those with long delay times live in dense regions within massive halos that have low star formation. We show that with ${\mathcal O}(10)$ events from current gravitational-wave detector networks, it is possible to make preliminary distinctions between formation channels which trace  stellar mass, halo mass, or star-formation rate. We also find that strategies to follow up gravitational-wave events with electromagnetic telescopes can be significantly optimized using the clustering properties of their hosts.

\end{abstract}

%\keywords{dark matter}


\section{Introduction}

The first gravitational-wave (GW) detection of a binary neutron star (BNS), GW170817, detected by advanced LIGO~\citep{2015CQGra..32g4001L} and Virgo~\citep{2015CQGra..32b4001A}, ushered in the era of GW multi-messenger astronomy~\citep{GW170817:discovery,GW170817:MMA}.
GW170817 was observed in a broad swath of the electromagnetic spectrum, including as a short gamma-ray burst (sGRB), X-ray/radio afterglow, and optical kilonova~\citep{GW170817:GRB,2017ApJ...848L..20M,2017ApJ...848L..21A,Coulter:2017,GW170817:DES,GW170817:MMA}.

The formation and evolutionary history of BNS systems are not well constrained~\citep[see e.g.][]{2018arXiv181210065B}. These systems are usually characterized by a delay time between the initial formation of the binary and its eventual merger. The  distribution of these delay times is expected to follow  the distribution of the major axes of their orbits ~\citep{Peters:1964}, and in the case of massive O/B stars that collapse to form compact objects binaries, a power-law distribution $dN/da\propto a^{\beta}$ is assumed \citep{Sana:2012px, Kobulnicky:2014ita}.  The merger time delay distribution is therefore also expected to follow a power-law distribution, $dN/dt \propto t^\alpha$, with some minimum time delay $t_d$. In general binary neutron stars are expected to have $\alpha=-1$, but if the binary goes through a common envelope phase the power-law dependence can be steeper and closer to $t^{-1.5}$ \citep{Dominik:2012kk,2018arXiv181210065B, Safarzadeh:2019}.

Within the next few years the network of GW detectors are likely to provide a statistical sample of tens to hundreds of BNS detections~\citep{2018LRR....21....3A}. We expect a substantial fraction of these BNS events to have electromagnetic counterparts and identified host galaxies. Statistical studies of the properties of the host galaxies will provide a new window into understanding these systems. 
For example a recent study by \cite{Safarzadeh:2019} proposes to use the stellar masses of BNS host galaxies to infer the time-delay distribution of neutron star mergers. In particular, they forecast the constraints on the minimum delay time $t_d$ and the slope $\alpha$ of the time-delay distribution that will be possible with future events. They find that it will take $\mathcal{O}(1,000)$ GW detections with identified host galaxies to constrain the parameter space of time-delay models. With third-generation GW detectors, such as Einstein telescope and Cosmic Explorer, the redshift distribution of detected events alone will provide tight constraints on the star-formation rate/time-delay distributions, without additional information from the host galaxies \citep{2019ApJ...886L...1V,Safarzadeh:2019pis}. In the meantime, it is instructive to explore the complete range of host galaxy properties that potentially correlate with binary evolution. In this work, we explore how the properties of galaxies that host BNS mergers depend on the BNS merger timescales. We show how a full range of galaxy observables can be used to constrain details of BNS evolution.

In addition to stellar mass, there are other host galaxy observables that contain information about the evolution of BNS systems through cosmic time. 
The rate of BNS mergers in a given galaxy is a convolution of the galaxy's entire history of star formation over its lifetime and the delay time between formation and merger. Galaxies reside in the centers of dark matter halos, and their evolutionary history is tied to the evolution of the dark matter halo in the cosmic web \citep[for a detailed review of the galaxy--halo connection see][]{Wechsler:2018pic}. Even at a fixed stellar mass, for example, star-forming and quiescent galaxies have significantly different evolutionary histories, affected by the environments of their parent dark matter halos. Adding information about the star-formation rate of a BNS host can therefore provide important insights into the underlying formation mechanism. Recent work by \cite{Artale:2019tfl} uses a population synthesis model for BNS mergers coupled with a hydrodynamic galaxy simulation to predict host galaxy properties, with a focus on stellar mass and star-formation rate. Their results suggest that while the BNS merger rate correlates most strongly with a galaxy's stellar mass, it also depends on the star-formation rate of the galaxy.

The  mass of the parent halo can also in principle be probed by a variety of mass proxies, including X-ray measurements of the virial temperature of the gas or the velocity dispersion of the satellites or stars associated with it. The net amount of baryonic matter available for a galaxy to form stars is directly related to the depth of the potential of its parent dark matter halo, leading to a correlation between a galaxy's luminosity and its halo mass. In general, high-mass halos tend to host more luminous galaxies \citep[e.g.][]{Kravtsov:2003sg, Vale:2004yt}; however, it has been shown that the slope of the stellar mass-halo mass  relation becomes significantly shallower at high halo masses \citep[e.g.][]{Behroozi:2012iw, Wechsler:2018pic, Kravtsov:2014sra}, indicating that galaxies with the same stellar mass can exist in a range of halo masses. Considering that quenching of star formation and various astrophysical feedback processes that control stellar evolution in a galaxy vary as a function of the host halo potential \citep{Silk:1997xw, Croton:2006ys, Bullock:2000wn, Hopkins:2011rj}, this implies different evolutionary channels for galaxies with similar luminosities but different halo masses. Therefore, in principle,  the host halo mass can provide complementary information to the stellar mass of a galaxy with regard to its growth history. Further, particularly at high stellar masses, where star formation appears to be quenched, the parent halo mass and environment provides additional information, independent from the star-formation rate, about the merger history of the galaxy itself.  

A galaxy's environment can provide additional information about the history of the galaxy. For example, galaxies that live in dense local environments like clusters or galaxy groups tend to be less star-forming and redder, while the fraction of blue, star-forming galaxies dominate the population in the field or low density environments \citep{Balogh:1997bw, Dressler:1980wq, Dressler:1984kh}. In principle, the large-scale environment of a galaxy can also play a role in its star-formation history. Since halos form hierarchically, assembling through mergers of smaller halos, the nature of halo mergers and their frequency are both dictated by the environment around the initial density peak.
The halo's merger history can also be expected to impact the history of the galaxy that resides within it. There are many ways to parameterize this environment; one relevant cosmological measure is the  density of matter in the vicinity of a galaxy at different scales, which is often measured as an abundance or number density of galaxies in the neighbourhood of a target galaxy. This is an observable that is measurable in most galaxy surveys and can in principle be used to constrain delay time distributions of binary systems.

 The galaxy in which a binary forms evolves in the time between formation and the binary's eventual merger. Correlations between properties of the host galaxy at binary formation, and the properties of the host galaxy when the binary merges, can therefore, in principle, be washed out for sufficiently long delay times \citep{Zevin:2019wun}.  However, the evolution of the galaxy can be inferred by studying the properties of host galaxies extensively in the full cosmological context. In this paper we study a set of observational properties that contain information about the evolution of the galaxy through cosmic time. To study the distribution of galaxy properties we use the Universe machine simulations \citep{Umachine}. These simulations populate dark matter halos in a cold dark matter (CDM), N-body simulation with galaxies. The connection between galaxies and their halos is based on a semi-empirical model that parameterizes the correlation between the star-formation rate of a galaxy and the properties of its parent halo including the potential well depth, redshift, and assembly history using an extensive set of observational constraints.  

GW170817 was accompanied by an sGRB, and there there is thus evidence that (at least some) sGRBs are expected to be produced by neutron star mergers.  Previous studies into the host galaxies of sGRBs~\citep{2007ApJ...664.1000B,2010ApJ...725.1202L,2013ApJ...776...18F, Prochaska:2005qf} have already provided important insights into the progenitors of BNS systems. A comparison between the host galaxy of GW170817 and the host galaxies of cosmological sGRBs was carried out in~\cite{2017ApJ...848L..23F}. Furthermore, understanding the connection between binaries and their hosts provides novel cosmological probes, such as inferring the cosmic star-formation history of the Universe~\citep{2018ApJ...863L..41F,2019ApJ...886L...1V}. Characterizing the binary--host connection can also provide an understanding of the systematic uncertainties involved in measuring the Hubble constant from standard sirens \citep{Schutz:1986gp, Holz:2005df, Dalal:2006qt, Abbott:2017xzu}, and similar studies have already been used to understand the connection between supernovae and their host galaxies \citep{Brout:2018jch}.
%\daniel{I tried to rephrase this. okay?}\susmita{ modified it slightly} 
%\risa reworded a little bit further...

Here we explore constraints on the time-delay distributions resulting from analyzing observable properties of the host galaxies to binary neutron star mergers. In particular we focus on the distributions of four observables: the stellar mass, specific star-formation rate, halo velocity dispersion (a proxy for halo mass), and the local density around the hosts of GW events.  For a range of time-delay models, we forward model the distribution of host galaxy properties using a simulated galaxy catalog.  In \S\ref{sec:method} we describe the simulations, and in \S\ref{sec:results} we describe our method and main results. 

\section{Simulated Galaxy Catalogs}
\label{sec:method}

Given the star-formation history of a galaxy and an assumed distribution of delay times between formation and merger, we can calculate the merger rate, $\mathcal{R}$, 
at redshift $z_f$:
\begin{equation}
        \mathcal{R}(z_f)=\lambda\int_{0}^{t(z_f)} \frac{dP}{dt} (t_f-t) \Psi_g(t)dt,
    \label{eqn:rate}
\end{equation}
where $\Psi_g(t)$ is the star formation rate of the galaxy at  time $t$, $dP/dt$ is the delay time distribution, and $\lambda$ is an efficiency factor. We consider a power-law distribution in merger times, such that $dP/dt\propto (t-t_d)^{-\alpha}$, where $t_d$ is the minimum delay time below which no BNS system can merge. We use $t(z)$ to convert between redshift and time assuming Planck cosmology \citep{Aghanim:2018eyx}.

The star-formation history of a galaxy depends on the galaxy properties, such as its halo mass, stellar mass, and local environment. Therefore, the observed merger rate in the Universe at a given time is the convolution of the delay time distribution and the %\maya{
star formation rate, where the star formation rate depends on various galaxy properties.
%}. 

To emulate the distribution of galaxy properties in a cosmological context we use the Universe Machine simulation. Universe Machine provides a galaxy catalog with galaxy stellar masses and star- formation histories extending from $z = 0$ to $z = 10$. We use the publicly available galaxy catalogs created using the Bolshoi--Planck simulation \citep{Klypin2011}, which is a dissipationless CDM-only $N$-body simulation of a $250$ Mpc $h^{-1}$ volume with $2048^3$ particles in a Planck Cosmology, $\Omega_m = 0.307$, $h = 0.7$.

To study the hosts of BNS events at $z=0$, we track the galaxy properties of every object in the catalog that has a halo mass greater than $1.35\times 10^{10} M_\odot h^{-1}$, corresponding to halos with more than 100 particles in the simulation. We use the merger trees from the {\sc Rockstar} halo finder~\citep{Rockstar} to reconstruct the galaxies' star formation histories across cosmic time. We use the main branch of the merger tree for every galaxy in the catalog, which tracks a galaxy's most massive progenitors in time. The galaxy merger history is sampled at 164 log-spaced points between $z=0$ and $z=10$. 

We focus on four observable properties of the galaxies: the stellar mass of the galaxy, $M_\ast$, the galaxy's specific star-formation rate, sSFR ($SFR/M_{\ast}$), the velocity dispersion of the parent halo, $\sigma_h$, and the ratio of the local and large-scale density, $\Delta_r$. The stellar mass and specific star-formation rates are provided by the simulated galaxy catalog. We use the specific star-formation rate because it traces the strength of the current burst of star-formation with respect to the galaxy's underlying stellar mass, it correlates strongly with the observed color and morphology of a galaxy.   \citep{Guzman:1997qv, Brinchmann:2003db}.

Meanwhile, we use the velocity dispersion of galaxies within the parent halo as an observable proxy for the total halo mass. This velocity dispersion is computed using the velocities of galaxies within a halo's virial radius. For satellite galaxies (i.e., galaxies that reside in subhalos) we use the velocity dispersion of the parent halo in which the subhalo resides. For galaxies that reside in low mass halos, where a sufficient number of subhalos are not resolved, we use the mass of the halo to compute the velocity dispersion. To compute the local density around a given galaxy, we measure the number density of neighbouring galaxies as a function of their 3D radius from the given galaxy. We only include neighbouring galaxies with stellar masses larger than  $M_\ast=10^9 M_\odot h^{-1}$.  We define the observable $\Delta_r$ as the ratio of density within $0.6$ Mpc$\,h^{-1}$  compared to the density within $5$ Mpc$\,h^{-1}$. The smaller scale is chosen to represent a galaxy's local environment, sensitive to whether or not it exists as a part of a group or a cluster of galaxies, while the larger radius is representative of its large scale environment or neighbourhood beyond its own parent halo. 

In the following sections we study the properties of the host galaxies to BNS mergers as a function of the time-delay model. We estimate the number of events needed to draw inferences about the evolutionary history of merging binaries. 

\section{Results}
\label{sec:results}

\subsection{Weighted distributions}

\begin{figure*}
	\centering
	\includegraphics[width=0.4\textwidth, trim=0in 0in 0in 0in]{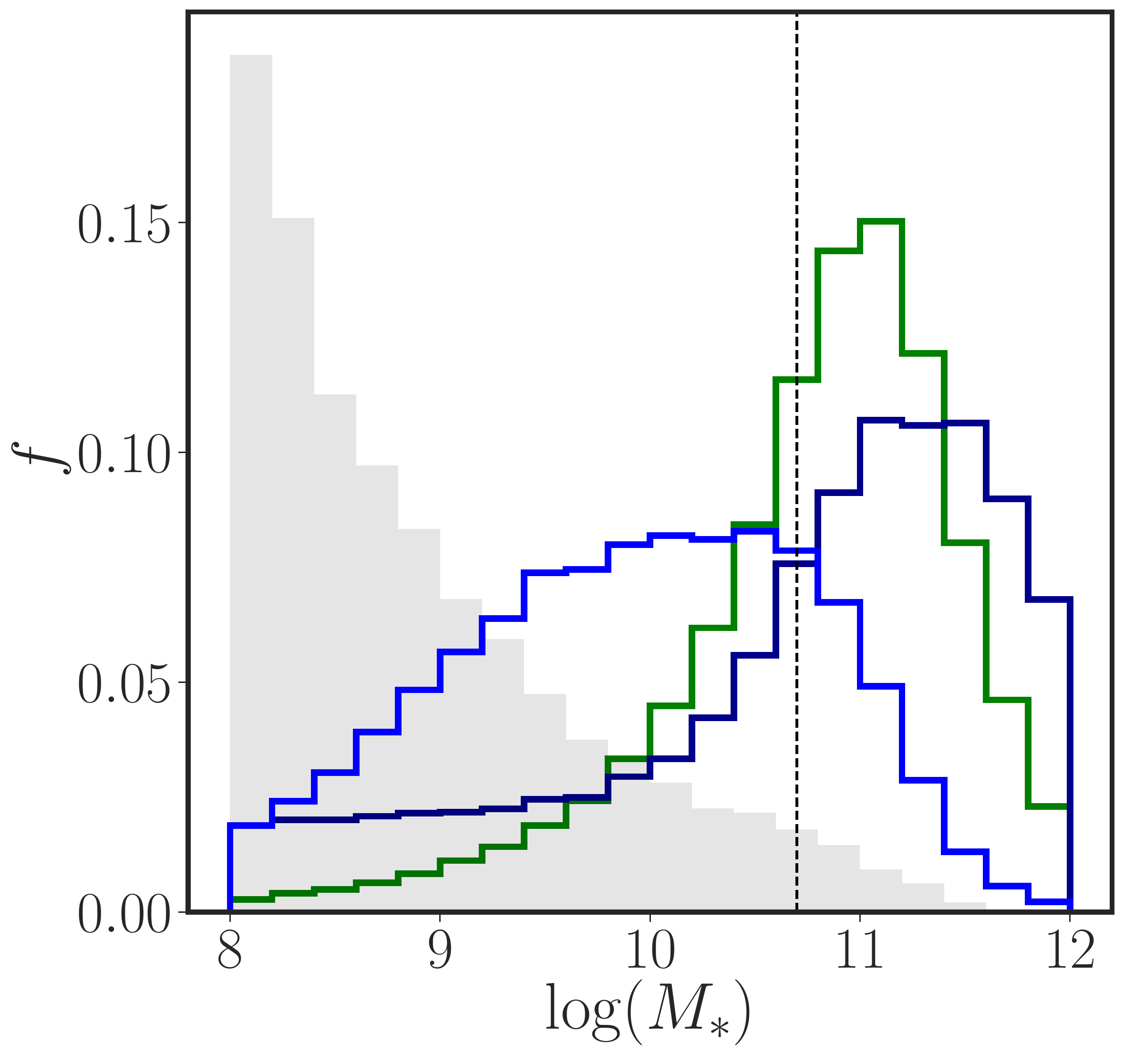}
	\hspace{0.2in}
	\includegraphics[width=0.4\textwidth, trim=0in 0in 0in 0in]{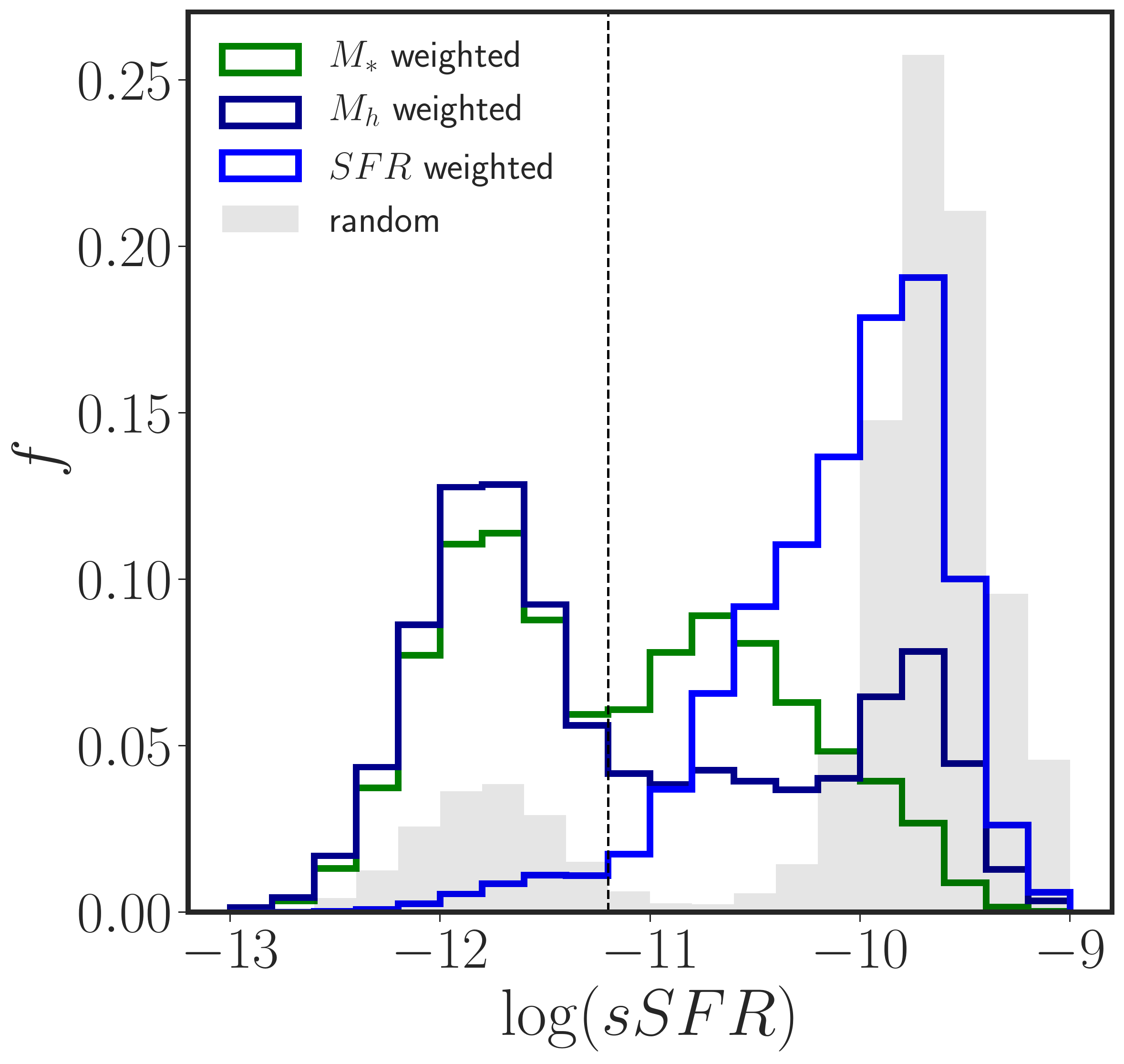}
	
	\caption{The distribution of galaxy stellar mass (left) and specific star formation rates (right) for the host galaxies corresponding to BNS events for models where those events are weighted by different galaxy properties.  We consider models in which BNS event rates are proportional to stellar mass, star formation rate, halo mass, or in which they are randomly assigned to galaxies above our mass threshold. The vertical dotted lines denote the stellar mass and specific star formation rate of NGC 4993. We note that the distributions show significant differences, suggesting that they might be distinguished with a small number of observations of host galaxies. }
	\label{fig:dists}
\end{figure*}

The underlying population of BNS systems is drawn from the distribution of stars in the Universe. Among the first questions that we can ask is whether BNS host galaxies are biased tracers of the underlying population of galaxies in the Universe. For example, we may ask whether BNS mergers trace stellar mass, in which case they would occur preferentially in high $M_\ast$ galaxies \citep{2019arXiv191105432D, Artale:2019tfl, Toffano:2019ekp}. Alternatively, BNS mergers may trace star-formation rate~\citep{1991ApJ...380L..17P} or halo mass, or they may be randomly distributed in galaxies with equal probability.

We ask how large a sample of host galaxies would be needed to distinguish these simple models where mergers trace either the stellar mass of a galaxy, the star formation rate, or the virial mass of the parent halo in which it resides. The arguments in this section are not specific to binary neutron star systems, and can be extrapolated to black hole binaries or black hole--neutron star mergers presuming their host galaxies could be identified.  

Physically, we can associate a model in which BNS mergers trace the star-formation rate to a zero delay time model, where binaries merge as soon as they are born. In this scenario the merger rate would simply trace the rate at which new stars are forming in a galaxy. At the other extreme, for very long time delays on the order of $10$ Gyr, we expect low-redshift BNS mergers to trace the stellar mass, as high stellar mass galaxies reached the peak of their star formation at early redshifts.
Alternatively, it has been proposed that rather than forming from isolated stellar binaries in the galactic field \citep{ Tauris:2017omb, Vigna-Gomez:2018dza}, a significant fraction of BNSs may form through dynamical interactions in dense globular clusters~\citep{Grindlay:2005ym, 2019ApJ...880L...8A}. 
Recent studies indicate that globular cluster abundance is a good tracer of a galaxy's host halo mass~\citep{Hudson:2014mva, Harris:2013zea}, and thus a BNS merger rate that traces dark matter halo mass rather than galaxy stellar mass could be an indication that BNS systems are preferentially formed in globular clusters. We also note that some studies suggest that formation in globular clusters is unlikely for BNS systems, but remains a viable scenario for BBHs, as massive black holes sink to the dense centers of globular clusters more efficiently due to mass segregation \citep{2019arXiv191010740Y}.

Figure~\ref{fig:dists} shows the distributions of stellar mass and specific star formation rates of host galaxies drawn from samples weighted by stellar mass, star formation rate, or halo mass. We choose a minimum stellar mass threshold of $10^8 M_\odot h^{-1}$ for this study, and only include galaxies above this threshold. The grey shaded region corresponds to a random selection of galaxies with (i.e. equal weights). 
We see that the  distributions are significantly different from a random sample, as well as from each other, and in principle even a small number of events can distinguish them.

\begin{figure*}
	\centering
	\includegraphics[width=0.39\textwidth, trim=0in 0in 0in 0in]{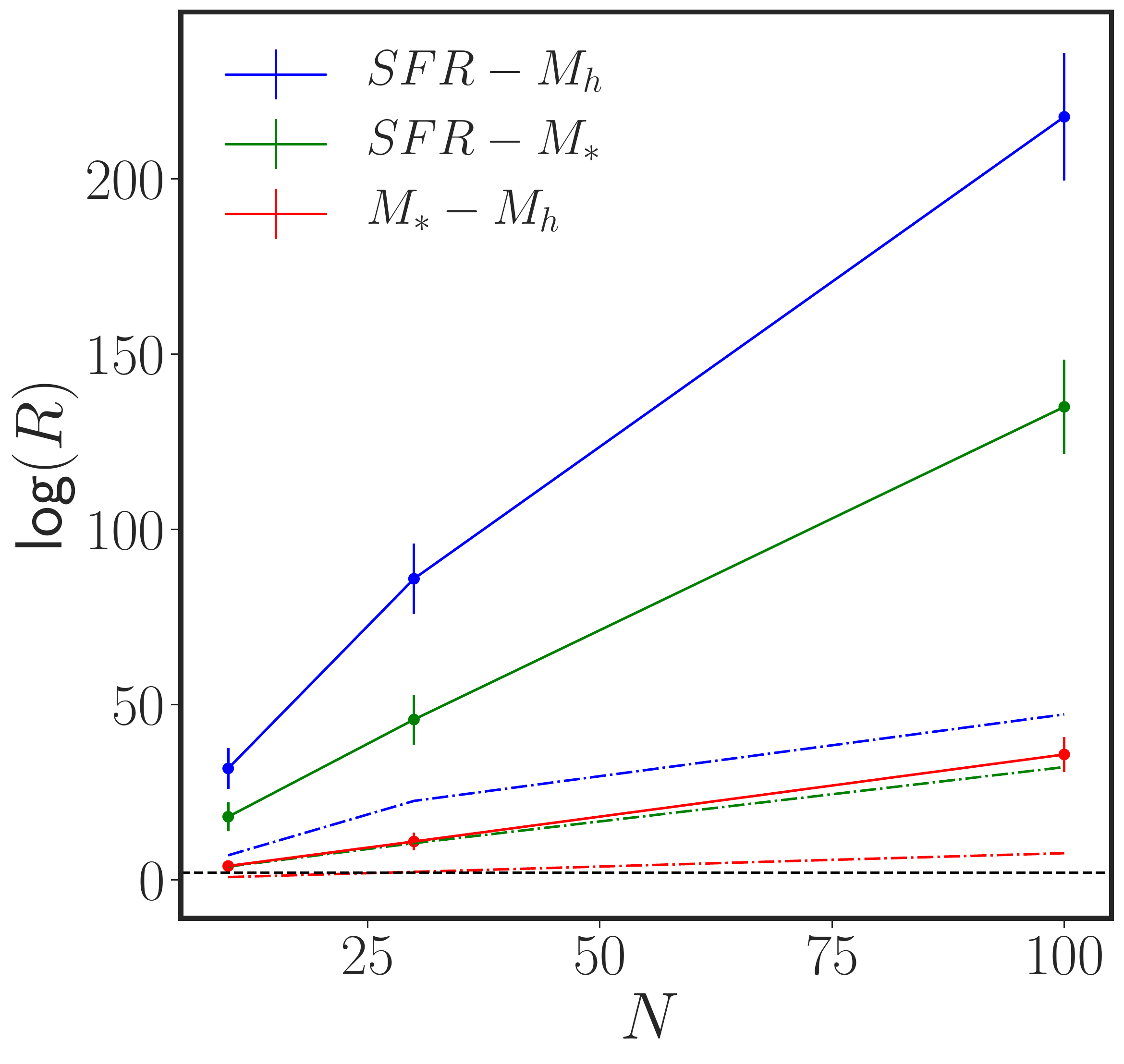}
	\hspace{0.8cm}
	\includegraphics[width=0.39\textwidth, trim=0in 0in 0in 0in]{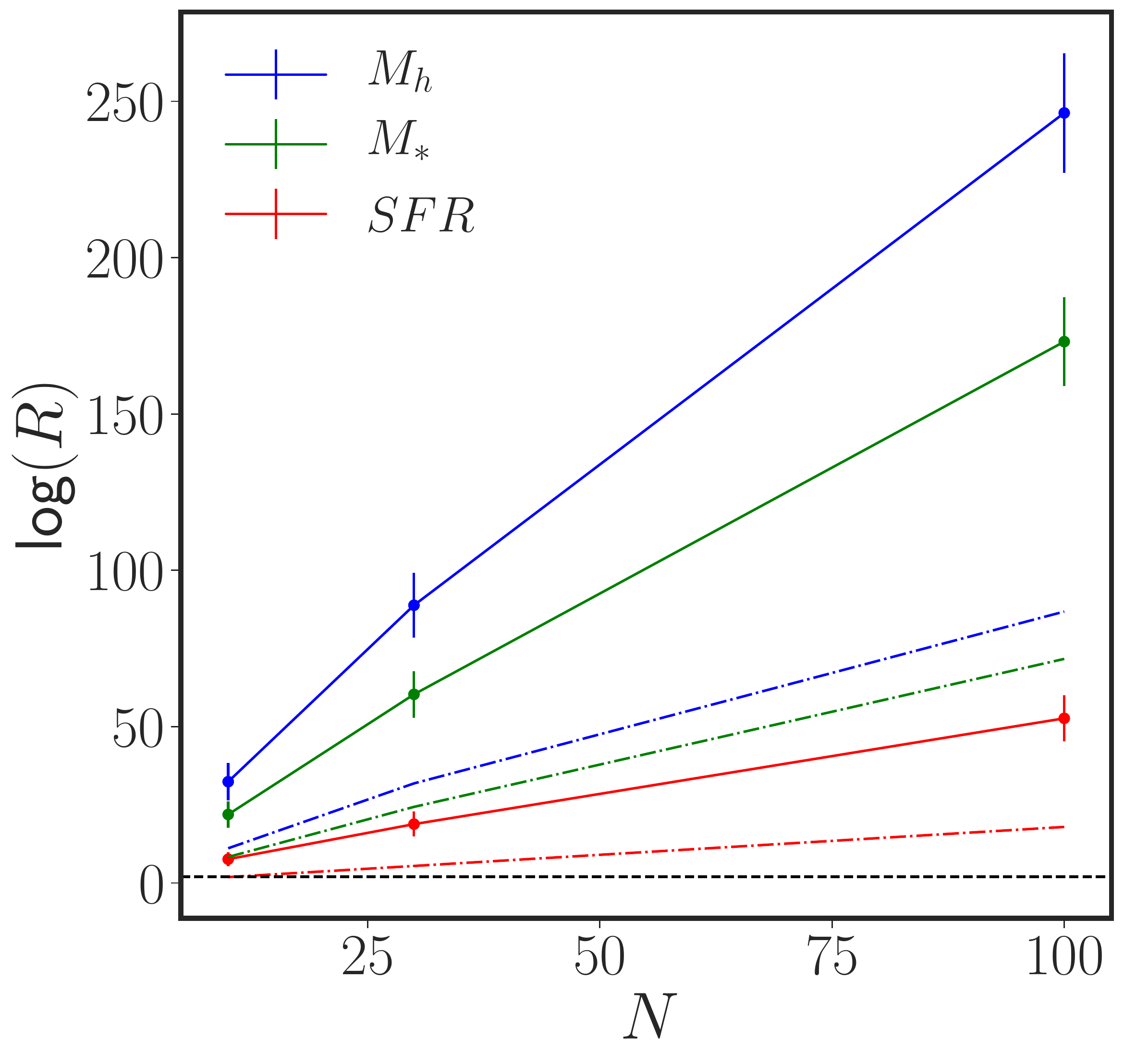}
	\caption{(Left) The logarithmic evidence ratio as a function of the number of events between models weighted by stellar mass and halo mass (red), stellar mass and star formation rate (blue), and halo mass and star formation rate (green). (Right) The evidence ratio between a random sample of galaxies and a weighted sample of galaxies. The dashed colored lines correspond to the evidence ratio when only the $M_\ast$ distribution of the host galaxies are used. The horizontal, dashed, black line corresponds to  a ratio of 99.7; points above this imply that models can be distinguished at greater than $3\sigma$ confidence. 
	}
	\label{fig:bayes}
\end{figure*}

To estimate the number of host galaxies required to distinguish between the $M_\ast$-, SFR-, and $M_{h}$-weighted distributions, we first construct the probability distribution of galaxies in the 4D space of observable properties, $\left [M_\ast, \mbox{sSFR}, \sigma_h, \Delta_r\right ]$. We draw $10^6$ galaxies from the full galaxy catalog with weights proportional to the various parameters. We approximate the 4D probability density as a histogram; i.e. we distribute the galaxies in bins of the observable properties and assume that the probability distribution function for model $k$,  $p_k(M_{\ast}, \mathrm{sSFR}, \sigma_h, \Delta_r)$, is piece-wise constant in each bin.
We use 10 bins in each direction\footnote{We find that our analysis is insensitive to the number of bins. We also test our results using kernel density estimation instead of histograms and find that our results are consistent.}. 

Given the probability distributions expected under each model, we draw 1,000 samples of $N$ galaxies from each model, as well as random selections of galaxies from the galaxy catalog.  We compute the likelihood of each model, $k$, given the events:

\begin{equation}
\mathcal{L}_k = \prod_{i=1}^N \left[p_k({M_{\ast}^i, \mathrm{sSFR}^i, \sigma_h^i, \Delta_r^i})\right],
\label{rat}
\end{equation}
where $p(...)$ is piecewise constant in the bins, and where $i$ runs over each event. Finally we compute the evidence ratio, $R=\mathcal{L}_1/\mathcal{L}_2$, between them to see if the models are distinguishable. 

Figure~\ref{fig:bayes} shows the log evidence ratio for pure samples drawn from different models. The right panel shows comparisons between the weighted models and a random draw of galaxies. We find that we can confidently tell apart a weighted sample from a random distribution with as few as 10 events when we use our complete set of observables (solid lines). The left panel shows the evidence ratio for distinguishing between the different models themselves. We find that we can distinguish a star-formation rate weighted sample from a stellar mass weighted one with order of 10 events, while we begin to distinguish between a stellar mass weighted and a halo-mass weighted distribution with about 30 events. \textit{We find that even with the small samples of neutron star mergers with counterparts expected to be identified in the near future using the current generation of GW detectors, we can draw important inferences about the formation mechanisms of these systems based on properties of their host galaxies.} In fact, we expect to be able to learn about the underlying models using only the distribution of observed stellar masses, as shown by the dashed lines in the same figure. Meanwhile, adding in additional information about the remaining properties increases the evidence in favor of the correct model. 
While we can begin to distinguish between the simple weighted populations using a few tens of events, in the next section we investigate how well we can resolve the time-delay distribution. 

\subsection{Delay time distributions}

As described in \S\ref{sec:method}, the rate of binary neutron star mergers in a given galaxy is a convolution of the delay time distribution of the BNS systems with the star formation rate of the galaxy over its history (Eq.~\ref{eqn:rate}). For an assumed time-delay distribution characterized by a slope $\alpha$ and a minimum delay time $t_d$, we forward model the merger rate at $z=0$ for every galaxy in our catalog using its star formation history. These star-formation histories are extracted from simulated merger trees constructed with the Universe Machine.

\begin{figure}[b]
	\centering
	\includegraphics[width=0.415\textwidth]{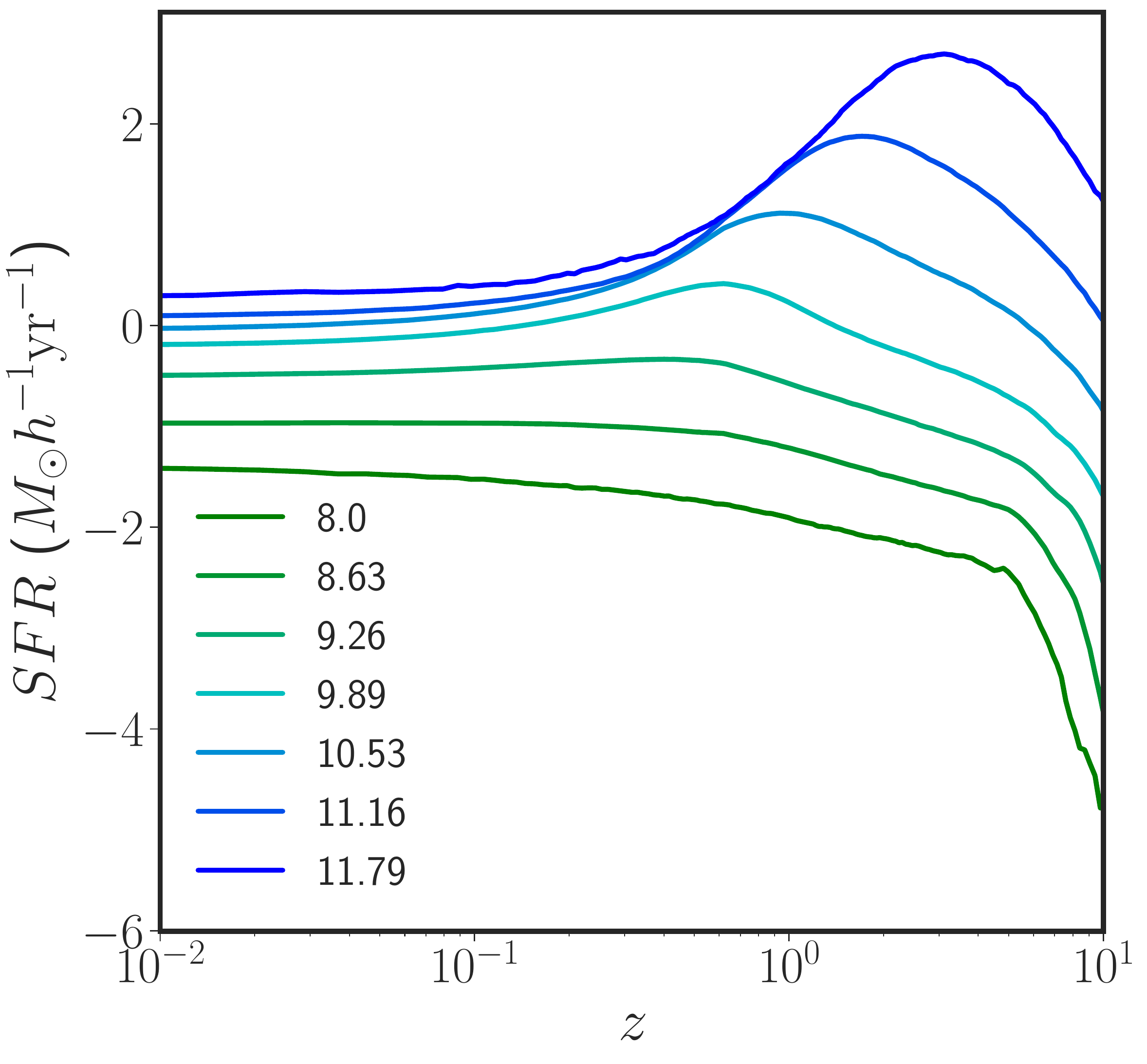}
	\caption{The average star formation histories  for different logarithmic stellar mass bins. The colors range from green to blue corresponding to low and high $z=0$ stellar masses in units of $M_\odot h^{-1}$. Galaxies with high stellar mass at $z=0$ reach the peak of their star formation at higher redshifts. 
	}
	\label{fig:sfh}
\end{figure}

Figure~\ref{fig:sfh} shows the average star formation history of galaxies of different stellar masses (where stellar mass refers to $z=0$). Today's massive galaxies tend to reach the peak of their star formation at earlier times compared to low mass galaxies. On the other hand, late time ($z\lesssim 0.5$) star formation histories are fairly flat, especially for low mass galaxies.
Star-forming and quiescent galaxies also tend to populate different regions of the Universe. In particular, galaxies in clustered environments today
are on average more quiescent than galaxies living in isolated environments. 
\begin{figure*}
\centering

\includegraphics[width=0.32\textwidth, trim=0in 0in 0in 0in]{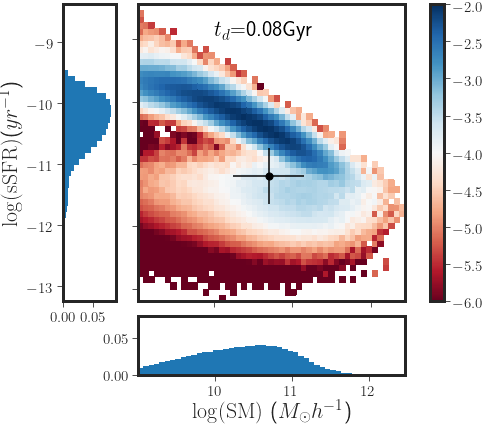}\hfill
\includegraphics[width=0.32\textwidth, trim=0in 0in 0in 0in]{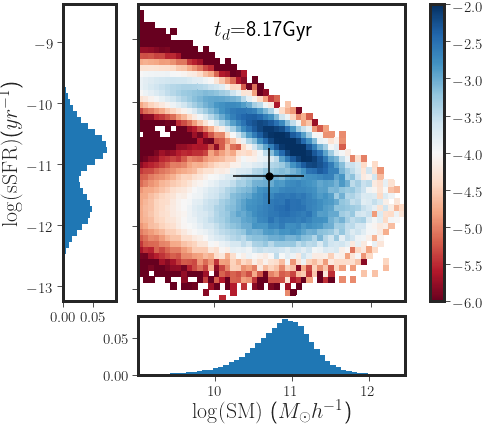}\hfill
\includegraphics[width=0.32\textwidth, trim=0in 0in 0in 0in]{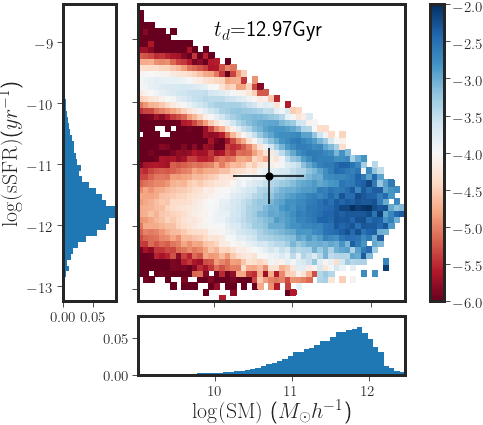}
\caption{Distribution of events in the sSFR-$M_\ast$ plane for host galaxies at $z=0$, for a delta-function delay time distribution with a time delay $t_d$. The three different panels from left to right are for subsequently longer time delays. The cross-bar is the position of NGC 4993.} 
\label{fig:time_evol_delta}
\end{figure*}

To build intuition as to how different delay times influence the properties of host galaxies, we consider an illustrative example in which the delay time distribution, $dP/dt$, is a delta function, $\delta(t-t_d)$. This corresponds to the scenario where all binaries merge instantly after a delay time, $t_d$, since their formation. For a given delay time the galaxies at  $z=0$  that are most likely to host a BNS merger are the ones that were forming the highest number of stars at the cosmological lookback time corresponding to the fixed $t_d$. Figure~\ref{fig:time_evol_delta} shows the probability distribution of GW events in the space of stellar mass and specific star formation rate for galaxies at $z=0$ for different, fixed delay times. The left panel shows the probability of a galaxy, with a given $M_\ast$ and $sSFR$ today, hosting mergers for a very short delay time. The dark blue band of highest probability follows galaxies that have the highest total star-formation rate (a product of their stellar mass and specific star formation rate) today. As the delay time increases from the left to right panels we see that the blue high-probability region shifts to galaxies with high stellar masses today. The extreme right panel corresponds to $t_d \sim 10\,\rm{Gyr}$. For such a long delay time, the BNSs merging in galaxies at $z=0$ were formed in galaxies at $z\sim 1$, which corresponds to the peak in star-formation for massive galaxies. Even though these massive galaxies have low star-formation rates today, they dominated the star formation in the Universe at $z=1$; as we shift from short to long time delays, the merger rate shifts from tracing SFR to stellar mass. Because the host galaxy of GW170817, NGC 4993, has relatively high stellar mass, but not a high star-formation rate, this single host galaxy mildly prefers long time delays.~\citep{2017ApJ...848L..28L,2017ApJ...848L..30P,2018arXiv181210065B, Blanchard:2017csd}.

\begin{figure*}
	\centering

	\includegraphics[width=0.38\textwidth]{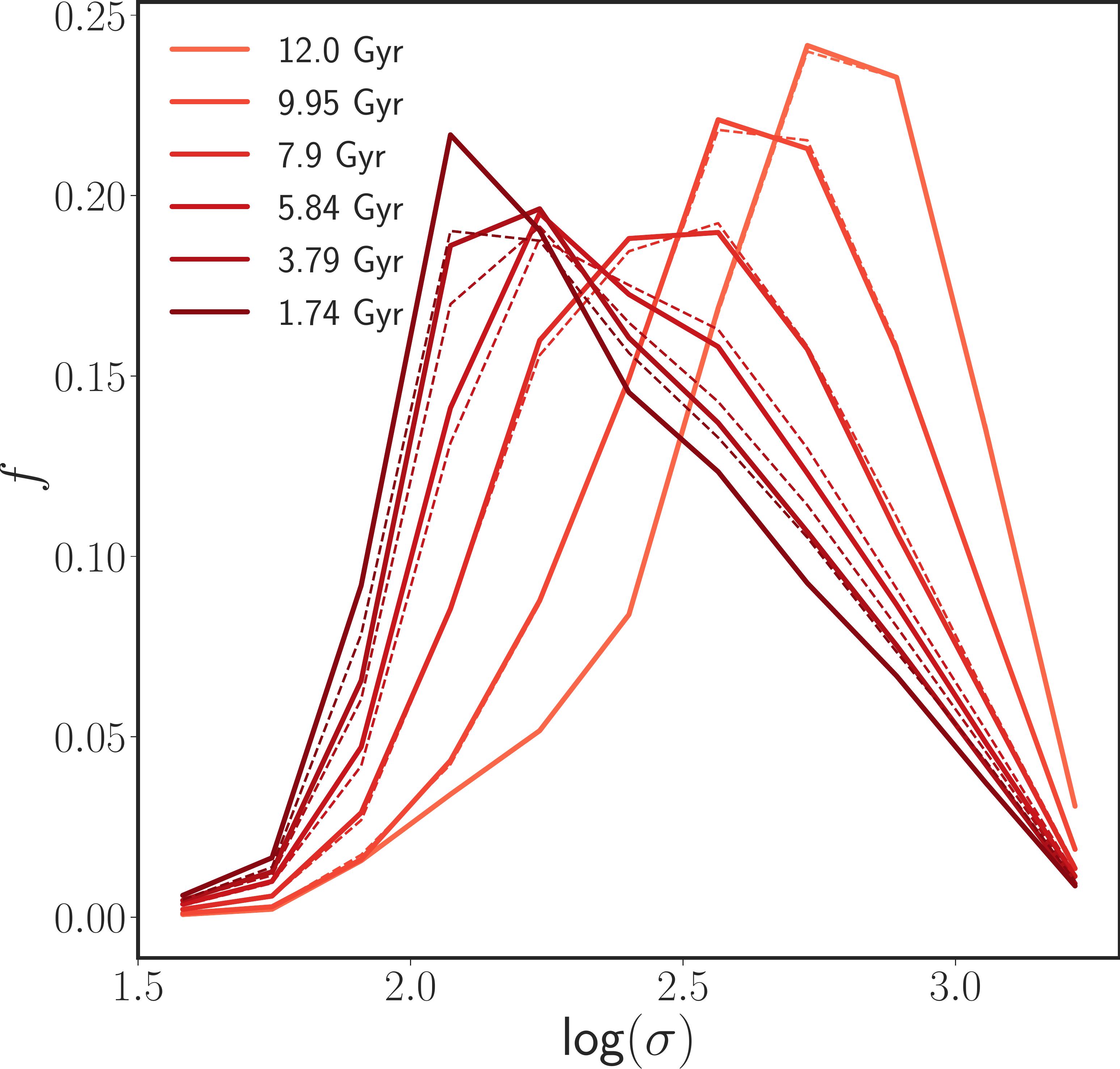}
	\hspace{0.5cm}
	\includegraphics[width=0.38\textwidth]{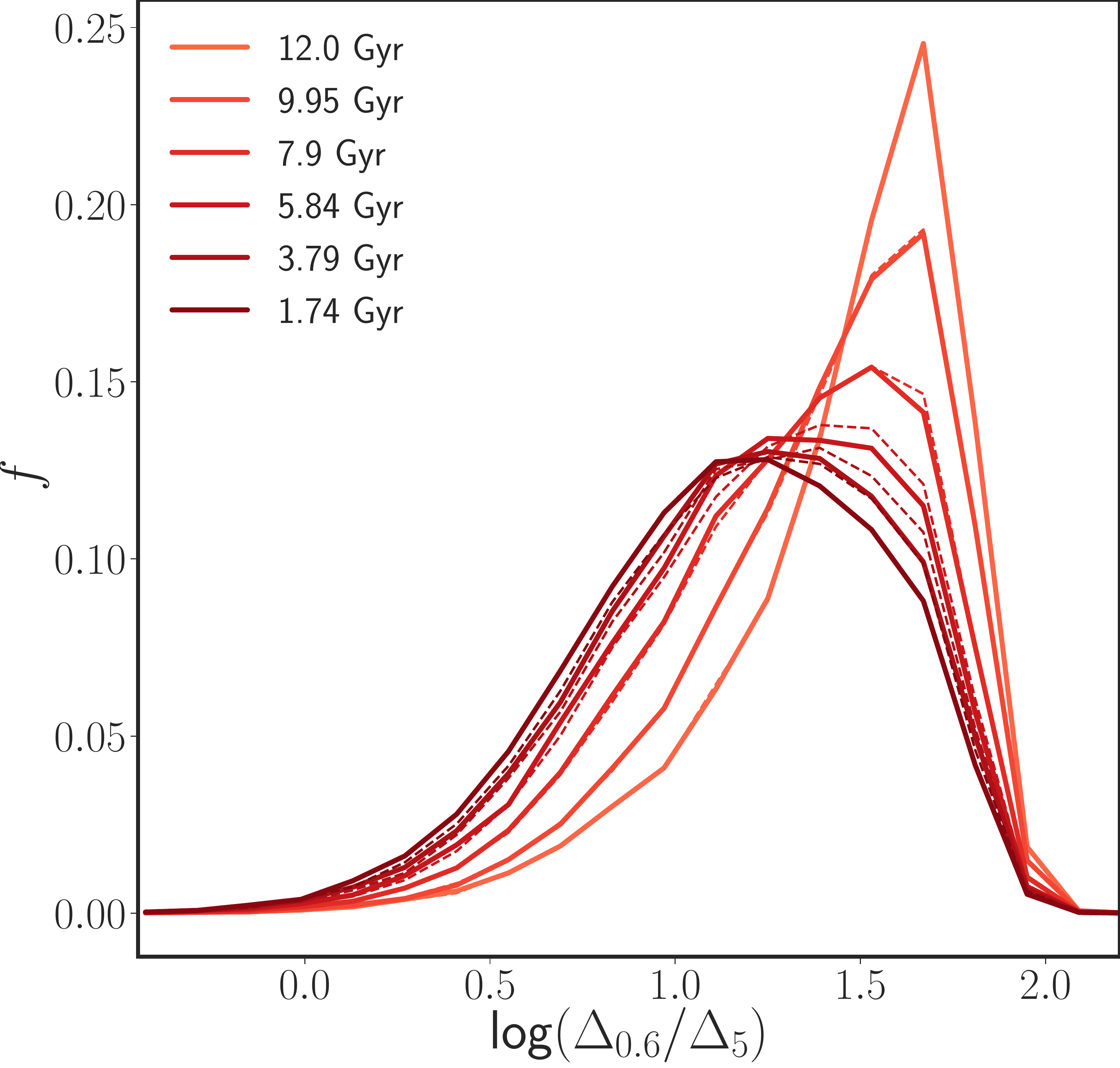}\\
	
	\caption{The distribution of velocity dispersions for parent halos of BNS mergers as a function of delay times. Velocity dispersions are a proxy for the parent halo mass, mergers with long delay times tend to live in higher mass halos. Right: Distribution of the ratio of enclosed densities, $\Delta_r$ withing $0.5\rm{Mpc} h^{-1}$ and  $5\rm{Mpc} h^{-1}$ around hosts of BNS mergers as a function of delay times. The solid lines correspond to $\alpha=-1$ while the dashed line corresponds to $\alpha=-1.5$, where $\alpha$ is the slope of the merger time distribution function.}
	\label{fig:halomass_dens}
\end{figure*}

While Figure~\ref{fig:time_evol_delta} demonstrates the distribution of host galaxies in the $M_\ast$--sSFR plane, Figure~\ref{fig:halomass_dens} shows the distribution of velocity dispersion (a proxy for halo mass) and density ratio (a measurement of the local environment density) of BNS hosts. We now consider a power-law distribution of delay times rather than a $\delta$-function, and vary the minimum delay time as well as the slope of the delay time distribution. 
For long delay times, BNS mergers preferentially occur in more massive halos, following the correlation between stellar mass and halo mass. This means that as the delay time is increased, the peak of the distribution of velocity dispersions in Figure~\ref{fig:halomass_dens} moves to larger values. We stress that for satellite galaxies (i.e. those that live in subhalos), the velocity dispersion refers to that of the parent halo in which the galaxy resides. The halo mass distribution is not particularly sensitive to the slope of the delay time distribution if it is varied between $-1$ and $-1.5$, except for fairly short minimum delay times. 

In the right panel of Figure~\ref{fig:halomass_dens}, we show the distribution of the local galaxy density $\Delta_r = \Delta_{0.6 {\rm Mpc} h^{-1}}/\Delta_{5{\rm Mpc}\,h^{-1}} $ surrounding BNS host galaxies at $z=0$. We measure the number density of galaxies with stellar mass greater than $10^{8} M_\odot h^{-1}$ enclosed within different spherical volumes around each host galaxy. The distribution of enclosed densities at large scales, of order $\sim 5~ {\rm Mpc}\,h^{-1}$, traces the overall bias of the halo while the enclosed density at smaller scales $\sim 0.5 {\rm Mpc}\,h^{-1}$, traces the local environment of a halo. For galaxies that reside in clusters or group-like environments, the small scale density should be higher. A high local density is correlated with high star-formation rate at earlier times. As star-formation histories are sensitive to the local environment, we find that the different delay time models lead to significantly different clustering properties among the host galaxies.

We now address how well we can expect to constrain the delay time distribution with a given number of BNS mergers with identified host galaxies. As in the case of the weighted models described above, under each time-delay distribution, we first construct the probability distribution of $z=0$ host galaxies in the observable space of $M_\ast$, $sSFR$, $\sigma_h$, and $\Delta_r$. We assume power-law time-delay distributions, and consider a wide range of minimum delay times, $0.1 < t_d < 12$ Gyr, and slopes,  $-4 < \alpha < 0$. 
Given $N$ host galaxy observations, we can use the forward-modeled distributions of galaxy properties to calculate the likelihood that the $N$ host galaxies came from a time-delay distribution with minimum delay $t_d$ and slope $\alpha$. We consider the case where we only have access to the stellar masses and star-formation rates of the observed host galaxies, as well as the case where the velocity dispersions and local densities are also available.

\begin{figure}[t]
	\centering
	\includegraphics[width=0.42\textwidth]{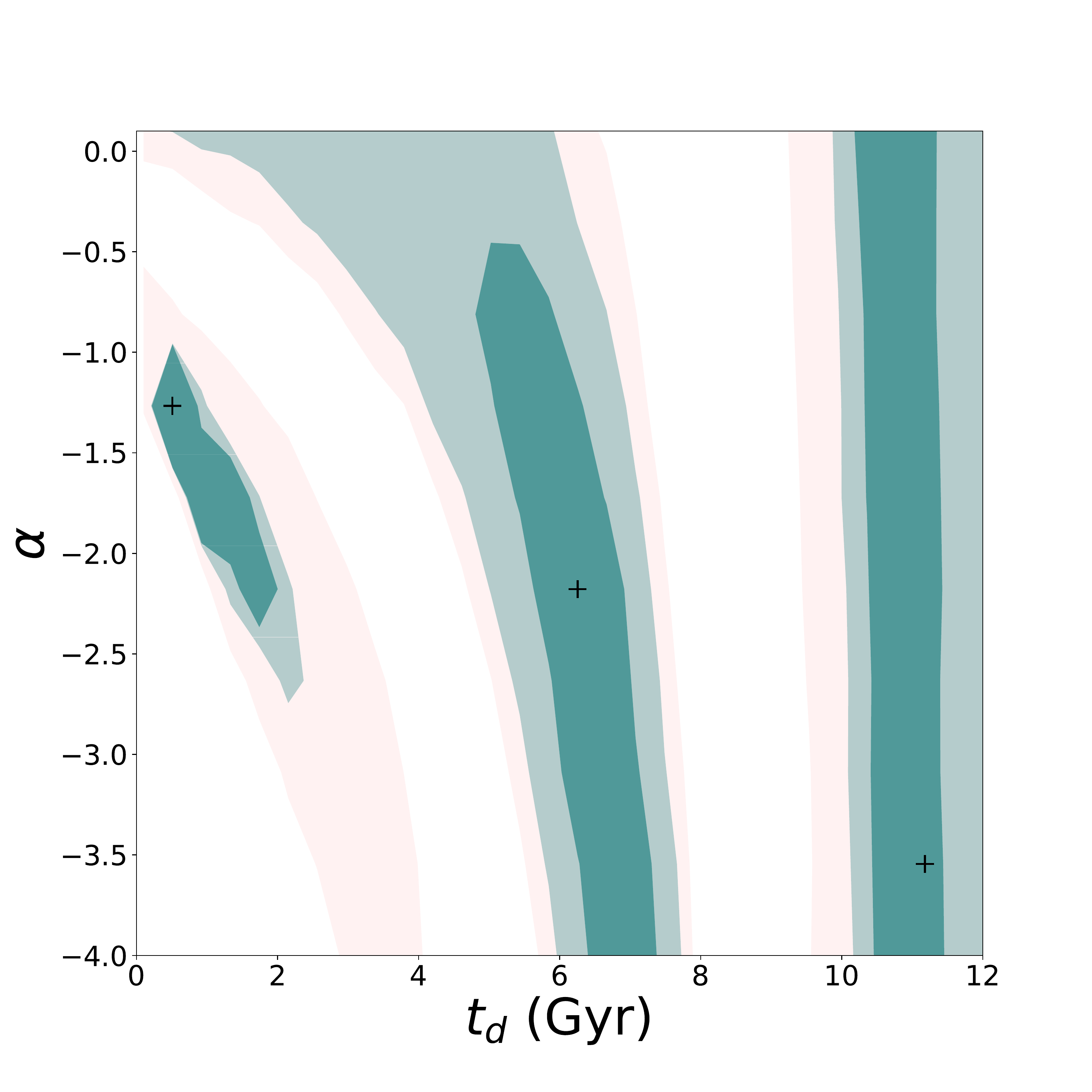}
	\caption{Posterior probability distribution of the slope and minimum delay of the time-delay distribution, as inferred for samples of 100 (dark green) and 30 (light green) host galaxies drawn from three different delay time models (crosses denote the truth) using the complete set of observables, [$M_\ast$, $sSFR$, $\sigma_h$, $\Delta_r$]. The pink shaded region corresponds to constraints obtained only from the stellar mass and the specific star-formation rate. We take flat priors on $\alpha$ and $t_d$, and so the posterior is proportional to the likelihood. The contours enclose the area under $68\%$ of the peak probability.}
	\label{fig:likelihood}
\end{figure}

In the coming years we expect tens to hundreds of BNS detections, and in Figure~\ref{fig:likelihood} we show the projected constraints from 30 and 100 events with identified host galaxies. We consider three different time-delay distributions with differing minimum delays and slopes, and show how well we can constrain each distribution. The shaded regions denote $1\sigma$ contours around each point. As a proof of concept, in this work we assume that the galaxy properties are perfectly measured (i.e. zero measurement uncertainty) and that each BNS merger is equally likely to have an identified host galaxy (i.e. there are no selection effects that bias host galaxy selection). The method can be easily extended to incorporate measurement uncertainty and selection effects if these prove to be significant.

We observe that using the velocity dispersion and the local density information of the host galaxies gives tighter constraints than using only the stellar mass and star formation rate information. We find that with $\sim100$ events we can constrain time delays with Gyr precision  and can confidently distinguish between low, moderate, and high minimum time delays. At small values of $t_d$, there is significant degeneracy between $\alpha$ and $t_d$. We also note that for large time delays (above $4$ Gyr) we lose our ability to constrain the slope of the merger time distributions (the contours are vertical).

\subsubsection{NGC 4993}

We assess our findings in the context of the BNS merger GW170817 and its host galaxy NGC 4993. GW170817 is currently the only BNS merger detected in GWs. Following the GW detection, electromagnetic searches of the GW localization volume succeeded in identifying its optical counterpart and associated host galaxy NGC 4993~\citep{Coulter:2017,GW170817:DES,GW170817:MMA}.
NGC 4993 has a stellar mass of $0.3$--$1.2\times 10^{11} M_\odot$ and an average star formation rate over a Gyr of $10^{-0.5} M_\odot \rm{yr}^{-1}$  \citep{Artale:2019doq}. Given its high stellar mass and low star-formation rate, it seems likely that NGC 4993 is drawn from a population that traces stellar mass more closely than star-formation rate (see Fig.~\ref{fig:dists}). Given this single host galaxy, we use its position on the $M_\ast$--sSFR plane to estimate the likelihood of different models. To account for the error bar in the measurement of $M_\ast$ and $sSFR$ we smooth the probability distributions while computing the likelihoods such that,
\begin{equation}
    \begin{split}
    p(M_\ast^{obs}, & sSFR^{obs} \mid H) \propto \\  & p(M_\ast^{obs}, sSFR^{obs} \mid M_\ast, sSFR)p(M_\ast, sSFR \mid H),
    \end{split}
\end{equation}
where $p(M_\ast^{obs}, sSFR^{obs} \mid M_\ast, sSFR)$ is a 2D Gaussian with given mean and standard deviation that corresponds to the $M_\ast$ and $sSFR$ of NGC 4993 and  $H$ is a model or a model and its parameters $\alpha$, $t_d$.

\begin{figure*}
    \centering 
    \includegraphics[width=0.78\textwidth]{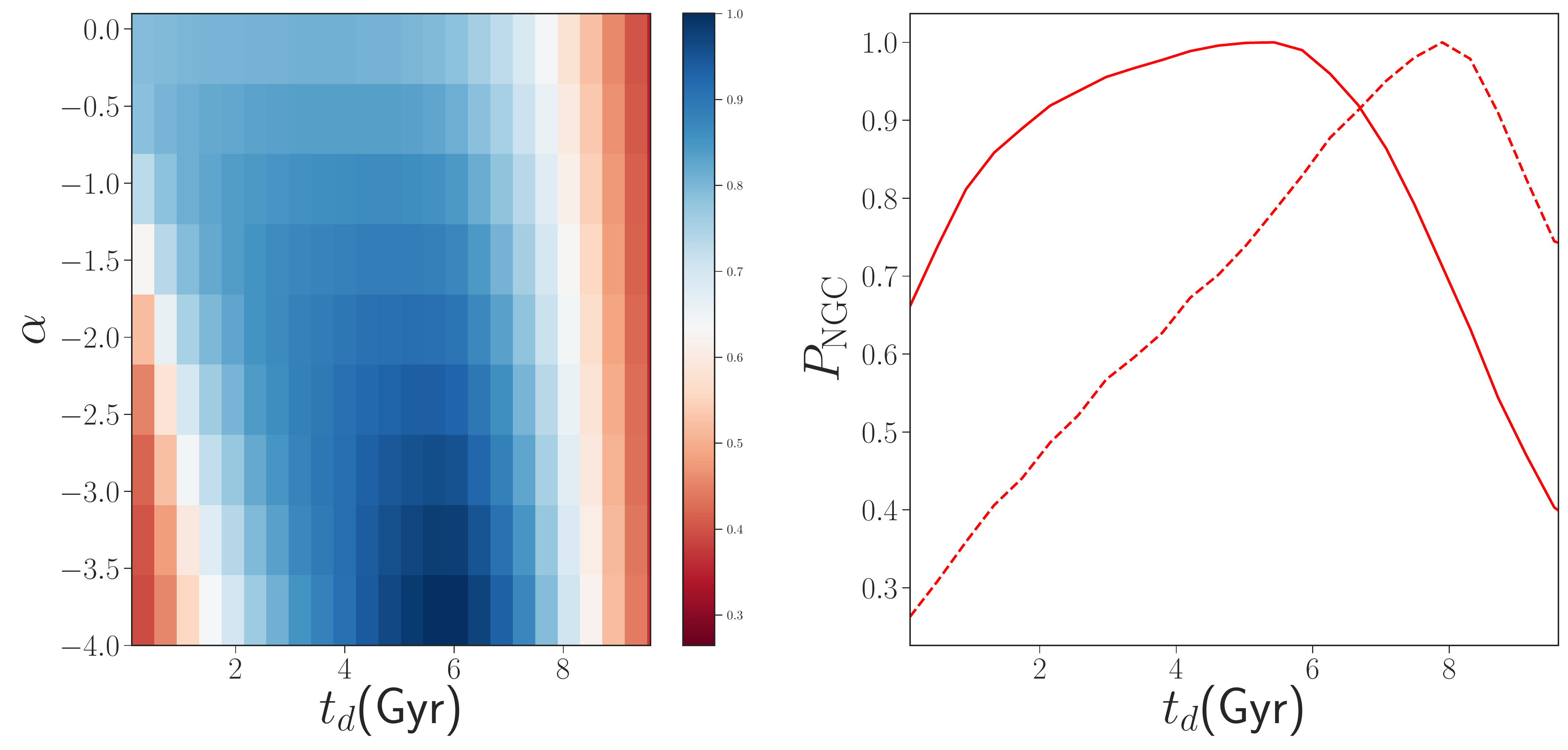}
    \caption{Probability of NGC 4993 hosting the merger event GW17087 for different merger time distributions with slope $\alpha$ and delay times $t_d$ with $SFR=0.5 M_\odot \rm{yr}^{-1}$ (averaged over a Gyr). The right hand panel shows the probability of a given delay time, marginalizing over all values of the slope. The dashed line corresponds to the probability with $SFR=0.01 M_{\odot} \rm{yr}^{-1}$. The probability distribution is sensitive to the value of star-formation rate. For the higher value of $SFR$ (solid), NGC 4993 shows a slight preference for an intermediate delay time, the lower value shows a preference for a delay time $\sim 8$ Gyr. }
    \label{fig:NGC_4993}
\end{figure*}

Figure~\ref{fig:NGC_4993} shows the probability of different time-delay distributions given the star formation rate and stellar mass of NGC 4993. We find that NGC 4993 prefers a time-delay distribution with a minimum delay time that is longer than $\sim 1 \mbox{Gyr}$ and a relatively steep slope. The right hand panel of Fig~\ref{fig:NGC_4993} shows the probability of different delay times when we marginalize over $\alpha$ with a flat prior in the range $-4<\alpha<0$. The preference for intermediate delay times is a consequence of the fact that NGC 4993 has a somewhat low average sSFR given its stellar mass. 

The value of $SFR$ quoted by \cite{Artale:2019tfl}, which corresponds to the solid curve in the right panel of \ref{fig:NGC_4993}, is the average star-formation rate over a Gyr, we note that NGC 4993 prefers an even longer delay time when we use the current star-formation rate of $0.01 {M}_\odot \rm{yr}^{-1}$ \cite{2018arXiv181210065B, Im:2017scv}. The dashed line in Fig. \ref{fig:NGC_4993} shows the probability of a delay time adopting the value in \cite{2018arXiv181210065B}, it appears that in this case,  NGC 4993 prefers a delay time that is closer to $\sim 8$ Gyr.   

Another aspect of NGC4993 that has been explored in the literature is the fact that it is not an isolated field galaxy but appears to be associated with a group of galaxies. \cite{Palmese:2017yhz} also found evidence for a dynamically driven formation for NGC 4993 from the existence of stellar streams in its photometry. As the number of GW events increase, it is of interest to ask how often the hosts of GW events tend to be satellite systems as opposed to isolated galaxies. We know from simulations that the fraction of satellite galaxies at a given stellar mass does not exceed $30\%$; comparing the satellite fraction of BNS host galaxies to the satellite fraction of all galaxies may be informative about the astrophysics of BNS systems. 

\begin{figure}[b]
    \centering

    \includegraphics[width=0.415\textwidth]{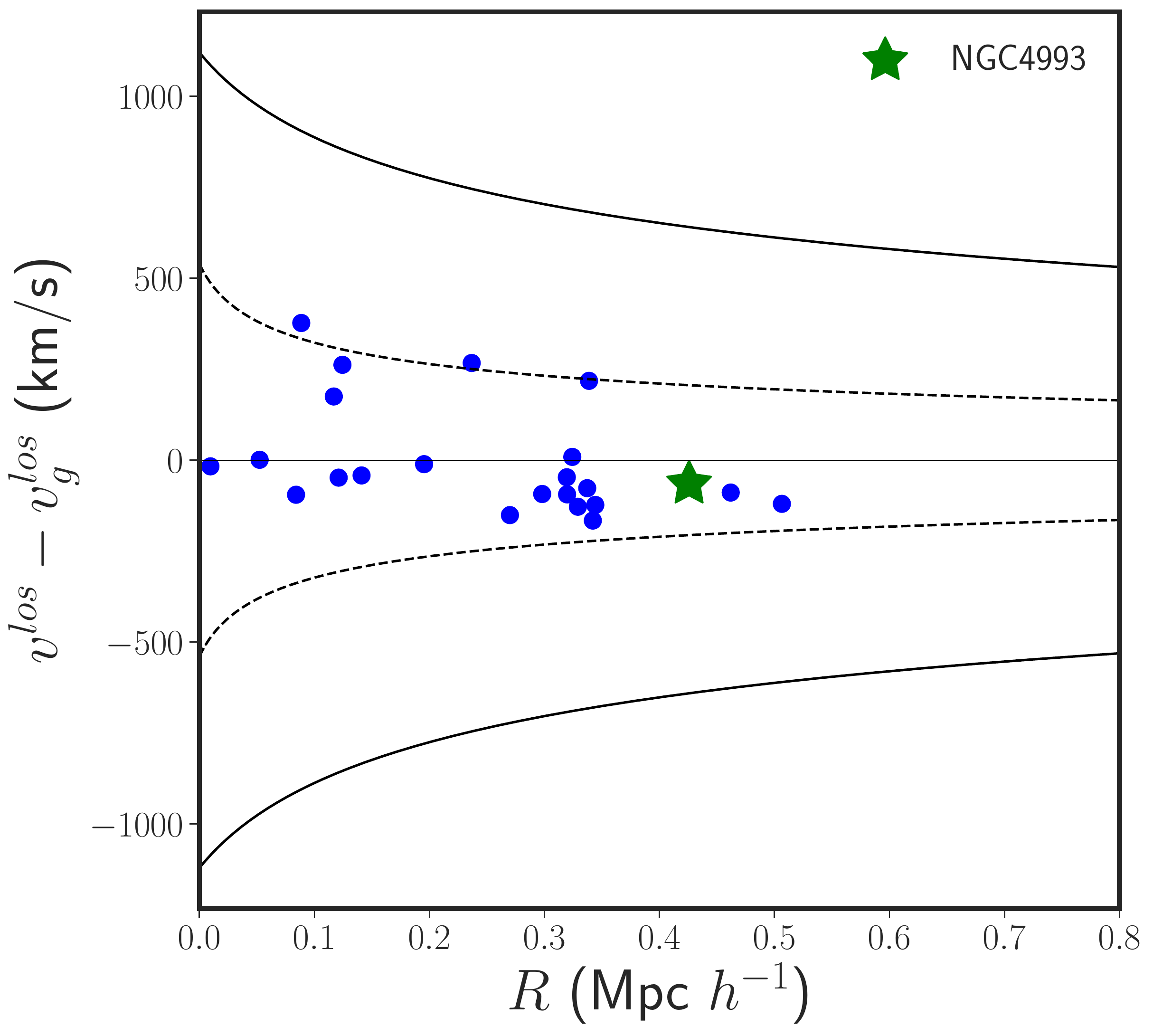}
    \caption{Projected phase space of galaxies in the group associated with NGC 4993. The black lines correspond to the escape velocity curves around a host halo of mass $10^{13} M_\odot h^{-1}$ (solid) and $10^{12} M_\odot h^{-1}$ (dashed).}
    \label{fig:NGC_phase}
\end{figure}

Table~2 in  \cite{Howlett:2019mdh} provides a  summary of the properties of the galaxy group that hosts NGC 4993 as a satellite. We note that it appears that there is significant inconsistency in the properties of the host group among the different works cited in the table. The measured velocity dispersion and richness of the group, for example, have a  large scatter. We also note that the most recent estimated virial radius of the group is $0.36{\rm Mpc}\,h^{-1}$  \citep{2017ApJ...843...16K}, which corresponds to a parent dark matter halo virial mass of $\sim 10^{13} M_\odot h^{-1}$); the velocity dispersion reported in the previous studies is too small for such an object. In Figure~\ref{fig:NGC_phase} we plot the projected phase space distribution of the member galaxies of the group that hosts NGC 4993 from \cite{2017ApJ...843...16K}. The escape velocity envelope (denoted by the black solid line) of a $10^{12} M_\odot$ galaxy is more representative of the population that is said to be associated with the group; such a halo has a virial radius of $\sim0.2 {\rm Mpc}\,h^{-1}$  which would place NGC 4993 in the outskirts of this group. Given evidence for a dynamically driven formation of the galaxy, it is possible that NGC 4993 is a galaxy within the splashback radius of the group. However, its high stellar mass also makes it less likely to be at a large halo centric distance of a group whose halo mass is as low as $10^{13} M_\odot$, as dynamical friction will draw it towards the center of the halo unless it is on first infall.

In summary, the star-formation rate and stellar mass of NGC 4993 already provide important insights into properties of BNSs. While the properties of the parent halo can add further information about BNS formation, it appears that further investigation is required to ascertain the parent halo mass of NGC 4993. As the sample of GW events increases, it will be useful to cross-correlate them with existing group and cluster catalogs, or follow up their host galaxies with spectroscopy to get the dynamical mass of the parent halo from neighboring galaxies to better understand the binary-host connection.

\begin{figure*}[b]
	\centering
	\includegraphics[width=0.39\textwidth]{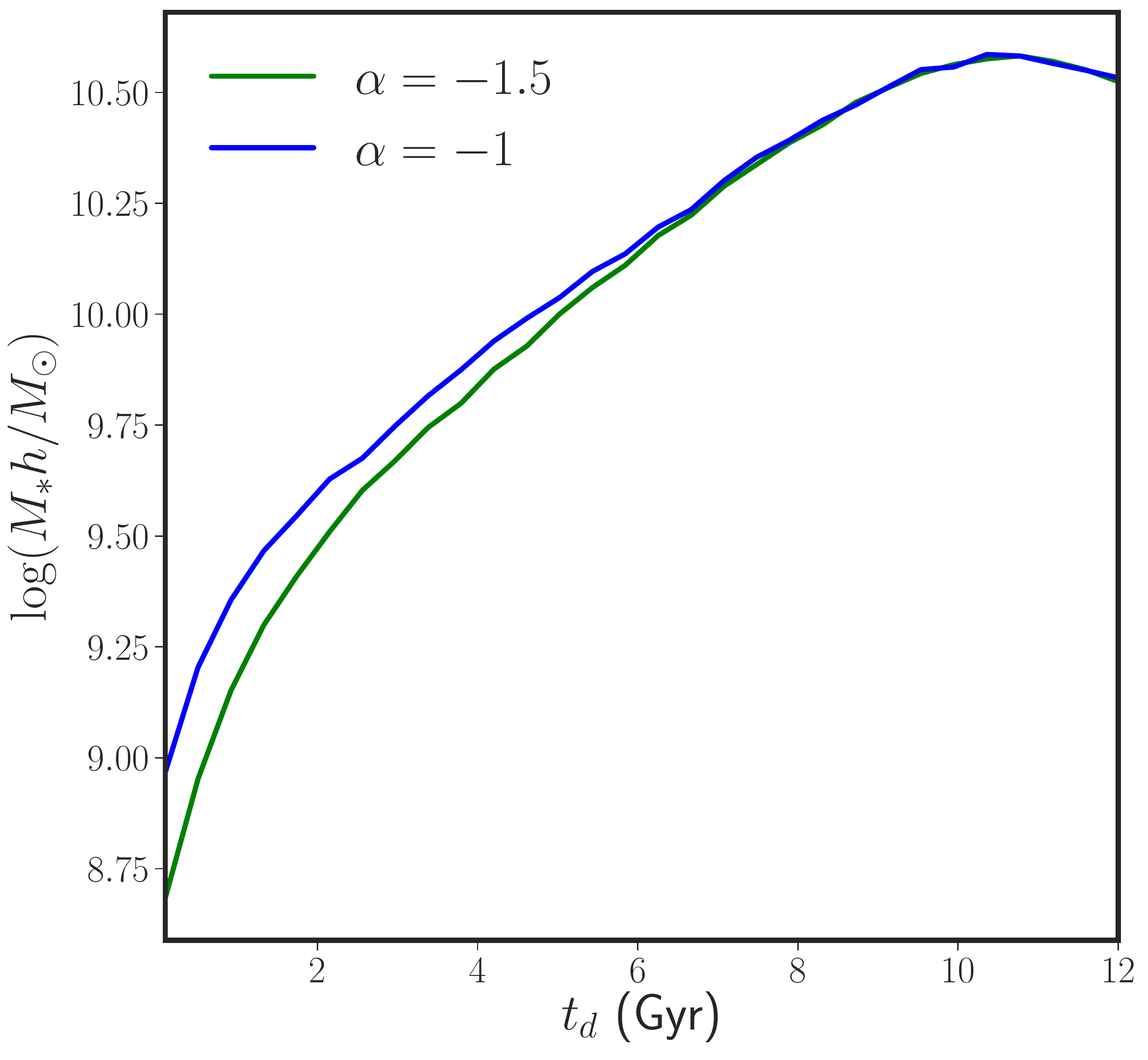}
	\vspace*{2ex}
	\includegraphics[width=0.38\textwidth]{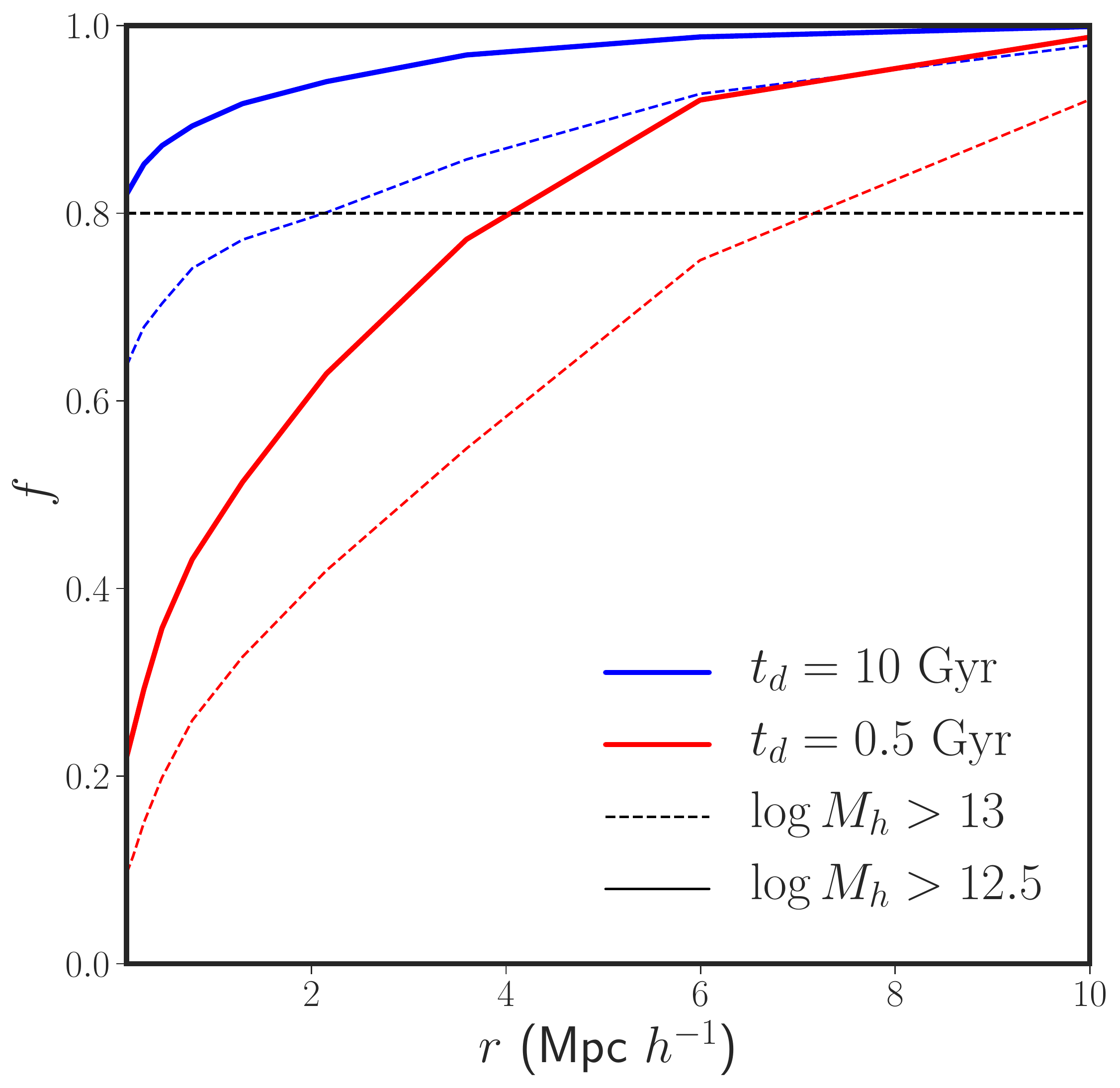}
	\caption{(Left) Stellar mass enclosing 90\% of potential counterparts of BNS gravitational wave events as a function of delay time. The different colors correspond to different slopes of merger time distributions, $\alpha$. (Right) The fraction of events that are enclosed with a radius $r$ of group mass halos above $M_h>10^{13} M_\odot h^{-1}$ (dotted lines) and $M_h>10^{12.5} M_\odot h^{-1}$ (solid lines), for two different time-delay models with minimum time delay $t_d = 10$ Gyr (blue) and $t_d = 0.5$ Gyr (red). These curves assume a slope of the time-delay distribution of $\alpha = -1.5$. These clustering statistics are calculated using a $250~{\rm Mpc}\,h^{-1}$(per side) simulation box.}
	\label{fig:minstar}
\end{figure*}

\subsection{Survey strategy}

As seen in the previous sections, different BNS formation models affect the population of their host galaxies. The previous sections were concerned with using observations of host galaxies in order to uncover the underlying formation models. However, in this section we discuss the inverse problem, and focus on how a known formation model can aid the observational follow-up effort and help identify the most likely host galaxies. In a three-detector network, the GW signal can typically localize a BNS source to a few tens of thousands of Mpc$^3$ on the sky, leading to hundreds of potential host galaxies in the 90\% credible volume~\citep{Chen:theone}. Many optical telescopes searching for a counterpart within a large volume rely on a pointing strategy to target the most probable galaxies first~\citep{2016ApJ...829L..15S,2019arXiv191105432D, Arcavi:2017vbi, Antolini:2016die}. Additionally, identifying the most probable galaxies in the volume (and assigning their relative probabilities of hosting a BNS merger) is integral to the galaxy-catalog based approach of measuring the Hubble constant~\citep{1986Natur.323..310S,2019ApJ...871L..13F}. 

The left panel of Figure~\ref{fig:minstar} shows, for different delay time distributions, the stellar mass threshold that would capture $90\%$ of all BNS events in the local Universe. As different delay time distributions trace the stellar content of the Universe in different ways, we see that the stellar mass threshold increases with the delay time of the merger. For the shortest delay times with a minimum time delay $\leq 1$ Gyr,  galaxy catalogs would need to go to a stellar mass depth of $M_\ast = 10^8  M_\odot h^{-1}$ to capture $90\%$ of BNS events. This corresponds to $\sim0.01 M_{\ast}^{\rm MW}$. A galaxy survey like DESI, which is complete to r-band magnitude $\sim19.5$, will be able to observe a complete sample at the stellar mass threshold out to $z=0.05$. 

While the depth of the survey is important for a follow-up strategy that searches through a galaxy catalog for a counterpart, another possible follow-up strategy may be to start with a cluster or group catalog and search the vicinity of the most massive halos. The success of such a strategy depends on the clustering of gravitational-wave events.
 
Massive halos that form from the collapse of rare peaks in the early Universe tend to cluster together. Galaxies that grow in the potential wells of halos also trace the clustering of their parent halos at all scales. Moreover, the stellar mass of a galaxy and the total dark matter content are correlated. Therefore, if the BNS merger rate traces stellar mass, BNS events are more likely to be clustered within the vicinity of the most massive objects in the Universe. Because the halo mass function falls off steeply at the high-mass end, searching for BNS counterparts around group-mass objects may significantly improve the efficiency of the search without requiring a deep galaxy catalog. 
  
However, the clustering of BNS host galaxies, just like other galaxy properties, is a function of the underlying delay time model. Under different models, the BNS merger rate traces the stellar mass of a galaxy in different ways. Additionally, the star formation history of a galaxy is environment-dependent. We find that for long delay times, BNS host galaxies are typically found within a smaller radius to a massive galaxy. The right panel of Figure~\ref{fig:minstar} shows the fraction of BNS events that are captured within spheres of different radii around the most massive halos. The solid and dashed curves correspond to halos with mass greater than $10^{13} M_\odot h^{-1}$ and $10^{12.5} M_\odot h^{-1}$, respectively. While for long delay times we can cover $\sim80\%$ of potential host galaxies by targeting the region within a few Mpc around the $10^{13} M_\odot h^{-1}$ groups, for shorter time delays, a group catalog that goes to lower halo masses would be more optimal.

\begin{figure*}
		\centering
	%\vspace{1cm}
	\includegraphics[width=0.9\textwidth, trim=0in 0in 0in 0in]{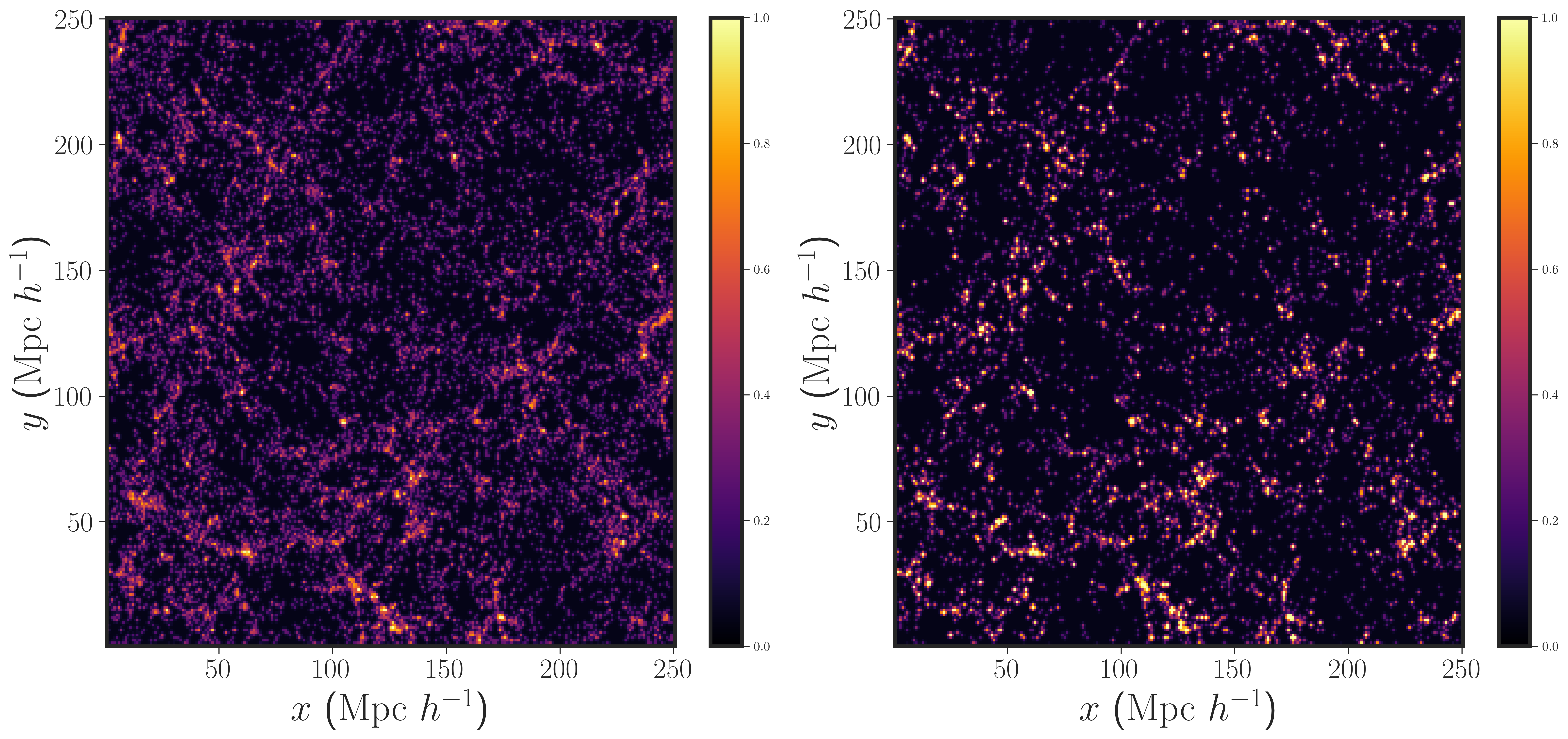}
	\caption{Distribution of GW events in space, taken from our simulations. The left panel corresponds to a delay time of $500$ Myr, while the right panel corresponds to a delay of $10$ Gyr. The color bar corresponds to the logarithmic density of events. The GW events are more clustered for long delay times (right) because most of these events are in the high stellar mass objects which dominated star formation at lookback time of $10$ Gyr. High $M_\ast$ galaxies live in high mass halos that tend to cluster more strongly. These plots assume a slope $\alpha = -1.5$ for the time-delay distribution.}
	\label{fig:clustering}
\end{figure*}

Figure~\ref{fig:clustering} shows the distribution of GW events in the simulation box over a $50 {\rm Mpc}\,h^{-1}$ slice. The left and right panel correspond to the distribution for a short and long delay time, respectively, bracketing our range of parameters. As we saw in Figure~\ref{fig:minstar}, GW events at $z=0$ are more clustered when they have long delay times compared to short ones. This is related to the fact that for long delay times, GW events trace galaxies with high stellar mass whose star-formation peaked at earlier times compared to low $M_\ast$ galaxies. The more massive galaxies tend to live in massive halos, which form out of the collapse of rarer peaks in the early Universe and tend to cluster together.

\section{Summary and Conclusion}

We have used the Universe Machine galaxy evolution model to predict the formation and merger of binary neutron stars over cosmological time. This allows us to relate astrophysical properties of the evolution of binaries detected by gravitational-wave networks to observable properties of their host galaxies. 

We estimate the rate of mergers for each galaxy at $z=0$ from merger-time models parameterized by the slope of the time-delay distribution, $\alpha$, and the minimum delay time, $t_d$, to make mock observations of GW events and study the properties of their hosts. Current generation gravitational-wave detectors, such as Advanced LIGO/Virgo and LIGO A+, are expected to detect tens to hundreds of BNS mergers with counterparts in the coming years. We show that the combination of events from current GW detectors with current and future galaxy surveys is a particularly promising avenue that will  allow us to draw important inferences about the underlying astrophysical population of binaries. The main findings of our work are:
\begin{itemize}
    \item[(i)] With a sample of  $\mathcal O(10)$ host galaxies to GW events, it is possible to distinguish formation models that predominantly trace stellar mass, star formation rate, or parent halo mass from each other and from a random selection of galaxies (see Fig.~\ref{fig:bayes}). We note that with $\sim30$ host galaxies it is possible to distinguish between halo-mass weighting and stellar-mass weighting, and thereby distinguish between associated formation channels of binary systems in globular clusters as opposed to galaxies.
	
	\item[(ii)] Using the distribution of observable galaxy properties (i.e. stellar mass, star formation rate, halo velocity dispersion, and local density) we find that with order of a few tens of events we can  distinguish short and long delay times at $\gtrsim3\sigma$ (see Fig.~\ref{fig:likelihood}). For longer delay times, we find an inability to constrain the slope of the merger time distributions. Adding in the halo velocity dispersion information significantly improves the constraints over those obtained using only stellar mass and sSFR.
    
    \item[(iii)] In the context of our  single existing GW counterpart event, GW17087, we find that the properties of the host NGC~4993 prefer a delay time that is longer than a Gyr (consistent with previous work; see \citealt{2017ApJ...848L..28L,2017ApJ...848L..30P,2018arXiv181210065B}). This can be attributed to the fact that NGC~4993 has a lower than expected star-formation rate for its measured stellar mass. Using the current star-formation rate we find that NGC 4993 shows a preference for a delay time close to $8$ Gyr. Related to this is the inference that this host galaxy appears to be drawn from a stellar-mass weighted or a halo-mass weighted sample. We also note that, while in the literature NGC~4993 has been associated with a bound group of galaxies, the measured velocity dispersion of the galaxies in the group suggest that the estimated virial radius of the group may not be representative of its true halo mass. The galaxies in the purported group appear to favor a $10^{12}{M_\odot} h^{-1}$ halo, with NGC~4993 at its outskirts. 
	
	\item[(iv)] In general we find that the probability of a galaxy hosting a merger today is a function of the delay time model, where BNSs with short delay times preferentially exist in star-forming galaxies that inhabit halos that are similar or lower in mass than the Milky Way, while mergers with longer delay times prefer more massive galaxies that on average populate groups or clusters and tend to be more clustered in space. 
	
    \item[(v)]Follow up strategies to localize electromagnetic counterparts for gravitational wave events can be optimized by searching for transients around massive galaxy groups (see Fig.~\ref{fig:minstar}). While the clustering of events  is related to the underlying delay-time distribution, we find that even for short delay times nearly $80\%$ of all events can be captured by searching within a few Mpc of halos of mass $M$ > $10^{12.5} M_\odot h^{-1}$. We also infer if transients are not found around the most massive objects in the Universe, this can be interpreted as favoring a model with short delay times. Although the study in this paper is limited to the low-redshift Universe, consisting of a $250^3$ Mpc$^3$ $h^{-3}$ volume at $z=0$, we note that clustering of events is an important probe of the underlying population of binary systems and may also provide a potentially useful tool to optimize localization strategies for future surveys. 
\end{itemize}

As galaxy surveys become more sophisticated, properties of galaxy hosts such as halo velocity dispersions and 3D clustering can be measured with greater accuracy using spectroscopic instruments like DESI and WFIRST \citep{DESI:2016, Spergel:2013tha}. Moreover, photometric surveys like DES, HSC, and LSST \citep{Abbott:2005bi, Aihara:2017paw, LSST09120201} can be used to measure clustering in projected space, which contains similar information about halo mass and environment. A systematic study of the properties of the host galaxies that include the properties of the parent halo and its spatial and temporal clustering as a function of delay-time models is a powerful tool to determine the underlying astrophysics of binary systems. In fact, we emphasize that folding in properties of the parent halo can provide important information to distinguish dynamical formation channels of binaries in dense stellar regions like globular clusters, the individual stars in these systems are not born as binaries and are not required to follow the power law delay time models. 

While GW17087 is the first gravitational wave associated with the confirmed merger of a binary neutron star, other astrophysical phenomena like sGRBs are also expected to be sourced by mergers of BNSs; therefore a similar study of hosts of sGRBs can be used to constrain delay-time models. In fact, GRB150101B, an sGRB similar to GW17087 and conjectured to be associated with a neutron-star merger \citep{Troja:2018ybt}, also has a very low star-formation rate and a stellar mass similar to NGC4993 \citep{Xie:2016rcc}.  Furthermore, in the scenarios where host galaxies can be identified, it will also be informative to study the distribution of other probes of the underlying formation mechanisms like offsets between the site of the merger and the host galaxy \citep{Voss:2003ep, Perna:2002su, Belczynski:2002bq, Zevin:2019wun} in relation to the properties of the hosts. For example, \cite{Kelley:2010qx} studied the distribution of binary events in the local universe by modelling natal velocity kicks that create such offsets in an N-body simulation. We also note that recent work has shown that a small fraction of BBH mergers (on order $\sim 1$ per year) can be localized well enough to allow the identification of a unique host galaxy, despite not having associated electromagnetic counterparts \citep{chen:2016tys}. Therefore a comprehensive follow up of the properties of the host galaxies, including its halo mass, environment along with its stellar mass and star-formation history can provide a window into the formation models of binary black hole systems as well. We note that \cite{Lamberts:2016txh}, \cite{Artale:2019tfl} and \cite{Artale:2019doq} have made predictions regarding the host galaxy properties of BBH mergers from population synthesis models.

The advent of gravitational wave astronomy provides a new observational probe of the physics of binary systems; the formation of galaxies and their co-evolution with cosmological structure, on the other hand, has been studied extensively over the years both theoretically and observationally through simulations and large scale galaxy surveys. We show that in the unique opportunity where we can identify a host galaxy for a gravitational wave event, the distribution of host properties combined with knowledge of galaxy formation can provide important information about the evolution of binaries through time. In the absence of a electromagnetic counterpart an understanding of the binary-host connection can help localization strategies for probable host galaxies. We emphasize that further studies of the connection between binaries and their hosts through the history of the Universe in an exciting prospect that can lead to a better understanding of the cosmological and astrophysical information we can extract from  gravitational wave sources.

\section{Acknowledgements}
We thank Tom Abel, Kirk Barrow, Peter Behroozi, Antonella Palmese, and Michael Zevin for useful discussions.  This research made use of computational resources at SLAC National Accelerator Laboratory, a U.S.\ Department of Energy Office, as well as the Sherlock
cluster at the Stanford Research Computing Center (SRCC); the authors thank support of the SLAC and SRCC teams.
This research was conducted in part at the Kavli Institute for Theoretical Physics, supported by NSF grant PHY-1748958.
%The authors gratefully acknowledge the Red-and-Black Chairs of the University of Chicago Physics Department for their their support throughout the preparation of this manuscript.
MF and DEH were supported by NSF grant PHY-1708081.
This work was partially supported by the Kavli Institute for Particle Physics and Cosmology at Stanford and SLAC and the Kavli Institute for Cosmological Physics at the University of Chicago through endowments from the Kavli Foundation. 
MF was supported by NSF Graduate Research Fellowship Program grant DGE-1746045.
DEH also gratefully acknowledges support from the Marion and Stuart Rice Award.

\bibliographystyle{yahapj}
\bibliography{references}

\acknowledgements

%----------------------------------------------------------------------------------------
%	Appendix A
%----------------------------------------------------------------------------------------



\end{document}